\def\year{2020}\relax
\documentclass[letterpaper]{article} 
\usepackage{aaai20}  
\usepackage{times}  
\usepackage{helvet} 
\usepackage{courier}  
\usepackage[hyphens]{url}  
\usepackage{graphicx} 
\usepackage{graphicx, subcaption}  
\usepackage{amsmath,amssymb}
\usepackage{tabulary}
\usepackage{longtable}
\usepackage{booktabs}
\makeatletter
\newcommand*{\rom}[1]{\expandafter\@slowromancap\romannumeral #1@}
\makeatother
\title{Mining Unfollow Behavior in Large-Scale Online Social Networks \\
        via Spatial-Temporal Interaction}
\author{
Haozhe Wu,\textsuperscript{\rm 1\thanks{Authors contributed equally to this work}}
Zhiyuan Hu,\textsuperscript{\rm 1\footnotemark[1]}
Jia Jia,\textsuperscript{\rm 1\thanks{Corresponding author: J. Jia (jjia@tsinghua.edu.cn)}}
Yaohua Bu,\textsuperscript{\rm 2}
Xiangnan He,\textsuperscript{\rm 3}
Tat-Seng Chua,\textsuperscript{\rm 4}\\
\textsuperscript{\rm 1}Department of Computer Science and Technology, Tsinghua University, Beijing 100084, China \\Beijing National Research Center for Information Science and Technology (BNRist) \\The Institute for Artificial Intelligence, Tsinghua University \\
\textsuperscript{\rm 2}Academy of Arts \& Design, Tsinghua University, Beijing, China\\
\textsuperscript{\rm 3}School of Information Science and Technology, University of Science and Technology of China\\
\textsuperscript{\rm 4}School of Computing, National University of Singapore\\
wuhz19@mails.tsinghua.edu.cn, zy-hu16@mails.tsinghua.edu.cn, jjia@.tsinghua.edu.cn
}
 
\begin{document}

\maketitle

%
\begin{abstract}
Online Social Networks~(OSNs) evolve through two pervasive behaviors: \textit{follow} and \textit{unfollow}, %
which respectively signify relationship creation and relationship dissolution. %
Researches on social network evolution mainly focus on the follow behavior, while the unfollow behavior has largely been ignored. %
Mining unfollow behavior is challenging because user's decision on unfollow is not only affected by the simple combination of user's attributes like informativeness and reciprocity, %
but also affected by the complex interaction among them. %
Meanwhile, prior datasets seldom contain sufficient records for inferring such complex interaction. %
To address these issues, we first construct a large-scale real-world Weibo\footnote{https://www.weibo.com/} dataset, %
which records detailed post content and relationship dynamics of 1.8 million Chinese users. %
Next, we define user's attributes as two categories: spatial attributes~(\textit{e.g.}, social role of user) and temporal attributes~(\textit{e.g.}, post content of user).
Leveraging the constructed dataset, we systematically study how the interaction effects between user's spatial and temporal attributes contribute to the unfollow behavior. %
Afterwards, we propose a novel unified model with heterogeneous information~(UMHI) for unfollow prediction. %
Specifically, our UMHI model: 1) captures user's spatial attributes through social network structure; %
2) infers user's temporal attributes through user-posted content and unfollow history; %
and 3) models the interaction between spatial and temporal attributes by the nonlinear MLP layers. %
Comprehensive evaluations on the constructed dataset demonstrate that the proposed UMHI model outperforms baseline methods by 16.44\% on average in terms of precision. %
In addition, factor analyses verify that both spatial attributes and temporal attributes are essential for mining unfollow behavior. %
\end{abstract}

\section{Introduction}
\begin{figure}[!t]
\centering
\includegraphics[width=.9\columnwidth]{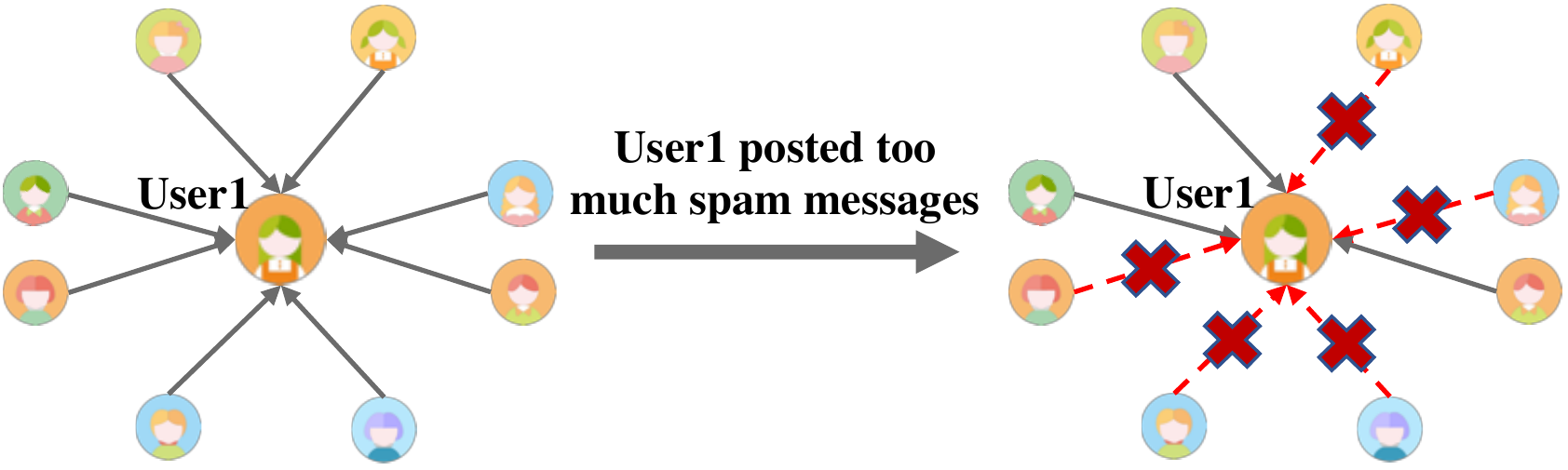}
\caption{An example of unfollow behavior in online social network: User1 posted too much spam messages, resulting the unfollow behaviors of his/her followers.}
\label{fig:Motivation}
\end{figure}

The popularization of the Internet greatly facilitates the development of Online Social Networks~(OSNs). %
Statistics\footnote{https://digitalreport.wearesocial.com/} show that more than 3 billion people around the world now use OSNs each month. %
The fast development of OSNs is tightly coupled with the evolution of online social relationships. %
There are two basic actions for users to manage their social relationships: \textit{follow}~(relationship creation) and \textit{unfollow}~(relationship dissolution), 
of which an example is shown in Figure~\ref{fig:Motivation}. %
Previous research efforts paid much attention to the follow behavior~\cite{liben2007link,bliss2014evolutionary,quercia2012tweetlda}, while unfollow behavior mining has largely been ignored. %
Statistics in a real-world Weibo dataset show that almost 40\% of users unfollow others at least once a month.
The frequent occurrence of unfollow behavior leads to an interesting question: why people unfollow others? %
Taking a step forward, can we predict the unfollow behavior in OSNs? %

Previous research efforts have shown the rationality of analyzing and predicting unfollow behavior. %
They mainly focus on mining unfollow behavior through user-posted content and social network structures, %
and have revealed several attributes that are closely related to the unfollow behavior, such as the informativeness of the followee and the reciprocity of relationships. %
Leveraging the found attributes, these methods predict unfollow behavior by defining handcrafted features~\cite{maity2018did,kwak2012more,xu2013structures}. %
However, merely defining handcrafted features from unfollow-related attributes cannot generalize well to large-scale setting, %
because user's decision on unfollow is not only affected by the simple combination of user's attributes, %
but also by the complex interaction effects among them. %
Take the community and post content of users as an example, %
user's personal habits shown in post content would influence the community of user, while the community would also affect user's preference on post content; %
thus the intertwined nature of community and post content would result in a highly complicated unfollow decision mechanism. %
Meanwhile, modeling such interaction effects needs abundant records of online social networks, %
while prior datasets seldom satisfy such requirement.

\begin{figure*}[!tp]
    \centering
    \includegraphics[width=2.0\columnwidth]{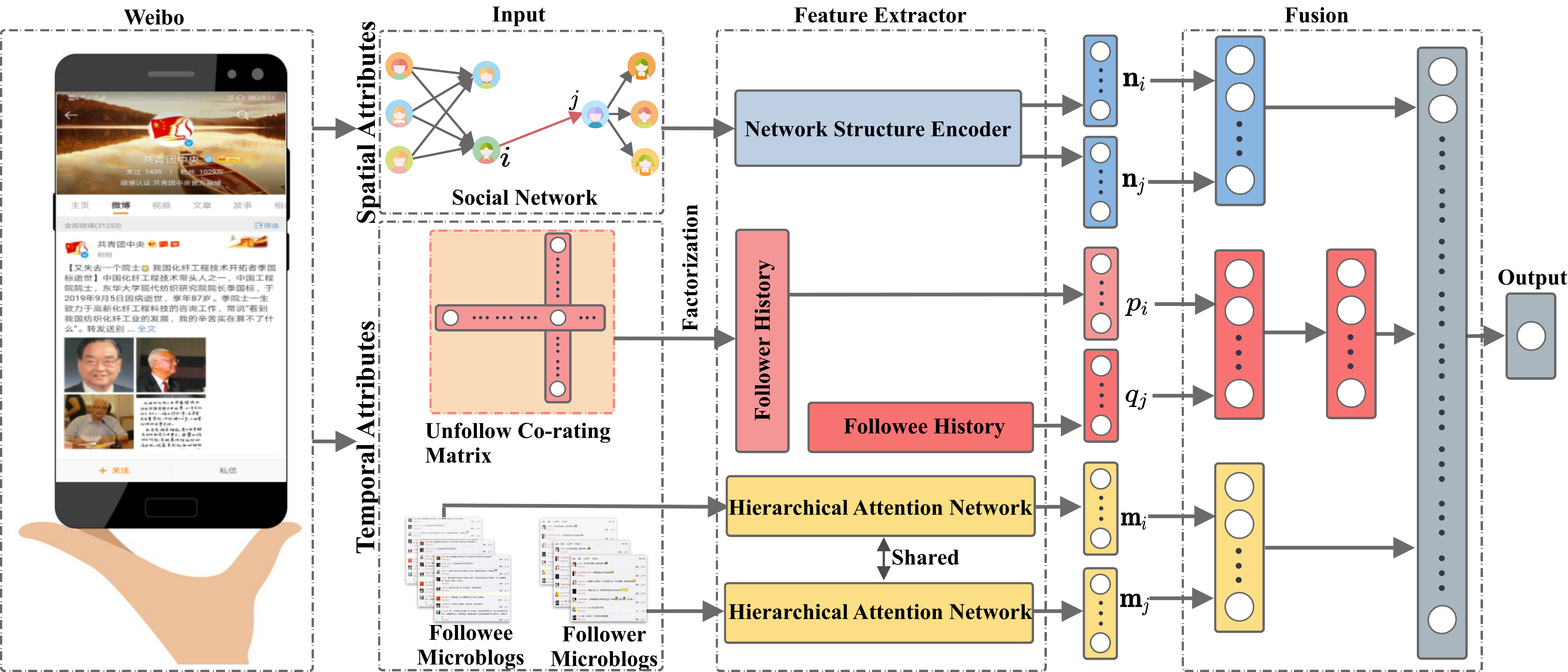}
    \caption{The workflow of UHMI: We predict unfollow behavior by fusing two kinds of information: (1) Spatial attributes, %
    here we utilize user-located network structure to represent such attributes. (2) Temporal attributes, %
    we incorporate user's temporal post content and unfollow history as the input to represent such attributes. %
    The interaction effects bewteen spatial and temporal attributes is modeled through the fusion stage in UMHI. %
    Details of hierarchical attention network is shown in Figure~\ref{fig:han}. } %
    \label{fig:workflow}
\end{figure*}

To address these issues, in this work, %
we first construct a large-scale benchmark dataset on Sina Weibo, %
which contains 1.8 million Chinese users, 400 million social relationships and 10 million records of unfollow. %
We record the timeline, content, and upvotes of each user's microblogs and track the unfollow actions of these users in a month. %
Then, inspired by previous researches, we define user's attributes as two categories: spatial attributes~(\textit{e.g.}, social role of user) and temporal attributes~(\textit{e.g.}, post content of user). %
Based on the constructed dataset, we systematically study how the interaction between spatial and temporal attributes contribute to the unfollow behavior and conduct exhaustive data observations. %
Next, for the unfollow prediction task, we propose a novel unified model with heterogeneous information~(UMHI) to learn the highly complex interaction. %
The main idea of UMHI is to model the user's spatial attributes through social network structure and user's temporal attributes through user-posted content and short-term unfollow history~~(a user's unfollowed-people list). %
First, information of social network structure is extracted by network embedding; %
Second, we adopt hierarchical attention network~(HAN)~\cite{yang2016hierarchical} to learn representations from user-posted content. %
Third, the matrix factorization~(MF) based collaborative filtering~\cite{koren2009matrix} is employed to reduce user's short-term unfollow history into low dimensional feature vectors. %
Finally, a unified heterogeneous information fusion network is trained to model the interaction between spatial and temporal attributes. %
Figure~\ref{fig:workflow} summarizes the workflow of our framework. %

Experiments demonstrate that our model outperforms the baseline methods by 16.44\% on average in terms of precision. %
In addition, factor analyses show that both spatial and temporal attributes are essential for mining unfollow behavior. %
To conclude, we summarize our contributions as follows:
\begin{itemize}
    \item We construct a real-world benchmark dataset on Sina Weibo with 1.8 million Chinese users and 400 million social relationships. %
    It records user's post content and relationship dynamics for a whole month. Such large-scale dataset is not only useful for unfollow prediction, %
    but also beneficial for further research like depression detection and rumor detection. %
    Dataset is publicly available at https://github.com/wuhaozhe/Unfollow-Prediction.
    \item We systematically study how the spatial and temporal attributes contribute to the unfollow behavior, %
    unveiling the interaction effects between these two categories of attributes. %
    \item We propose a novel UMHI model, which predicts unfollow behavior by learning spatial and temporal attributes through user's footprint on OSN. %
    The proposed method outperforms baseline methods by a large margin. %
\end{itemize}

\section{Related Work}
Previous researches on social network evolution mainly focus on the follow behavior~(link prediction and friend recommendation), while the unfollow behavior has received less scrutiny. %

For mining follow actions, several researches focus on predicting relationship from social network structures~\cite{tang2015line,perozzi2014deepwalk,grover2016node2vec}.
Perozzi \textit{et al}. proposed Deepwalk algorithm~\cite{perozzi2014deepwalk} to represent the network structure as low dimensional embeddings through random walks on social networks, %
then the similarity between embeddings reflects the possibility of establishing relations. %
The node2vec algorithm~\cite{grover2016node2vec} extends the depth first random walk strategy into a biased random walk procedure, %
making the learned representations scalable. The LINE algorithm~\cite{tang2015line} embeds node into low dimensional representations by optimizing the first order and second order proximity. %
All of these approaches only capture spatial factors, while unfollow behavior is caused by intricate interactions of spatial and temporal factors. %
Therefore the aforementioned algorithms all suffer from inferior performance in terms of unfollow prediction. %

Compared with these link prediction methods, researches on unfollow behavior mostly resort to rule based methods.  %
Kwak \textit{et al}. firstly researched on the unfollow dynamics in Twitter, they found some unfollow factors, including informativeness, reciprocity and relationship stabilization \cite{kwak2011fragile}. %
Later, they built a logistic regression model based on structure properties and behavioral properties \cite{kwak2012more}. %
The same group then adopted actor-oriented model~(SIENA) to examine the impacts of reciprocity, status, embeddedness, homophily, and informativeness on tie dissolution \cite{xu2013structures}. %
Kivran-Swaine \textit{et al}. explored how network structures alone influence unfollow behavior \cite{kivran2011impact}. %
Quercia \textit{et al}. researched on whether user's demographics such as age, gender will influence unfollow behavior in facebook \cite{quercia2012loosing}. %
Maity \textit{et al}. analyzed the content of the posts made by the Twitter users who lose followers consistently and extracted various behavioral features from followee's post content to make prediction. \cite{maity2018did}. %
However, these rule-based methods can hardly represent the interaction between the spatial and temporal attributes, therefore can't generalize well to large-scale settings. %

\section{Problem Formulation}
\label{sec:def}
The problem setting of unfollow prediction is to predict a user's future unfollow behavior from raw online social networks~(OSNs) data. %
OSNs contain several attributes that are predictive for unfollow behavior, we define them as two categories: spatial attributes and temporal attributes. %
Formally:
\begin{itemize}
    \item \textbf{Spatial Attributes}: user's attributes that remain unchanged in time interval $[t_{start}, t_{end}]$~(\textit{e.g.}, user's social role).
    \item \textbf{Temporal Attributes}: user's attributes that dynamically change in time interval $[t_{start}, t_{end}]$~(\textit{e.g.}, user-posted content).
\end{itemize}
Given a set of users $V$, we use a binary matrix $R \in \mathbb{R}^{|V|\times|V|}$ %
to denote the link dynamics between users in time interval $[t_{start}, t_{end}]$. Specifically, each entry $r_{ij}$ denotes whether user $i$ %
has unfollowed user $j$ in $[t_{start}, t_{end}]$, which is defined as:

\begin{equation}
    r_{ij}=\left\{
\begin{array}{lr}
    1, \quad \mbox{if user $i$ unfollowed user $j$ in $[t_{start}, t_{end}]$;} \\
    0, \quad \mbox{otherwise.}
\end{array}
\right.
\end{equation}

To conduct training-test scheme, given test edges $E_{test}$, we mask $E_{test}$ from binary matrix $R$. %
Specifically, for any $r_{ij} \in E_{test}$, we enforce $r_{ij} = 0$ to get the training binary matrix $R_{train} = R \backslash E_{test}$, %
here we call matrix $R_{train}$ to be the unfollow history matrix, %
$r_{i\cdot}$ to be the unfollow history of user $i$, and $r_{\cdot j}$ to be the unfollowed history of user $j$. %

\textbf{Problem.} The target of unfollow prediction is to predict $r_{ij}\in E_{test}$. We incorporate social network structure information $\mathbf{n}_{i}, \mathbf{n}_{j}$ as spatial attributes, %
incorporate posted content $\mathbf{m}_{i}, \mathbf{m}_{j}$ and unfollow history $r_{i\cdot}, r_{\cdot j}$ as temporal attributes. %
Then, our objective is to learn a function $y_{ij}=f(\mathbf{n}_{i}, \mathbf{n}_{j}, \mathbf{m}_{i}, \mathbf{m}_{j}, r_{i\cdot}, r_{\cdot j})$, %
which estimates the probability that user $i$ would unfollow user $j$.

\section{Dataset Observation}
\label{sec:data_obs}
In this section, we conduct exhaustive data observations to analyze how spatial attributes and temporal attributes interact with each other. %
In order to visualize the observation results, we define several statistics that respectively represent spatial attributes and temporal attributes. %
We take user's social role as an example of spatial attributes. Specifically, %
following the prior work of Yang~\cite{yang2016social}, we divide users into three groups: %
5\% of users with the highest PageRank score~\cite{page1999pagerank} are considered to be the opinion leader~(OpnLdr); %
5\% of users with the lowest Burts Constrain score~\cite{burt2017structural} are considered to be the bridges between disconnected communities in social network, \textit{a.k.a.}, the structure hole~(StrHole); %
and the rest of users are considered to be ordinary users~(OrdUsr). %
Then for the temporal attributes, we define the following two attributes: 
\begin{itemize}
    \setlength{\topmargin}{0pt}
    \setlength{\itemsep}{0em}
    \setlength{\parskip}{0pt}
    \setlength{\parsep}{0pt}
    \item Similarity: the tf-idf similarity between follower and followees' post content within a month.
    \item Exposure: the number of microblogs posted by the followee within a month.
\end{itemize}

Before analyzing the results of data observations, we first elaborate the details of dataset construction. %
We build a large-scale benchmark dataset on Sina Weibo, which contains 1.8 million Chinese users and 400 million social relationships. %
We record each user's microblog post content and relationship dynamics from September 28, 2012 to October 29, 2012. %
Each post is recorded with its time, content and upvotes. %
During the month we observe, we found that 10,705,319~(2.53\%) edges have been broken at least once, and 714,945~(40.00\%) users have unfollowed others at least once, 
verifying that unfollow is a pervasive behavior. 
\begin{table}[htbp]
    \centering
    \caption{Sum of edges in the $E_{test}$ among different social roles and relations statuses. OrdUsr, OpnLdr and StrHole are the shorthands of ordinary user, opinion leader and structure hole, %
    meaning that the followee of the edge is OrdUsr/OpnLdr/StrHole. }
    \begin{tabular}{lrrrr}
    \toprule
    Social role & OrdUsr & OpnLdr & StrHole & Sum   \\ 
    \midrule
    hold        & 1887          & 2551           & 1364           & 5802  \\ 
    unfollow    & 2391          & 2539           & 1860           & 6790  \\ 
    \midrule
    Sum         & 4287          & 5090           & 3224           & 12592 \\ \bottomrule
    \end{tabular}
    \label{tab:dataset_compose}
\end{table}

\begin{figure}[!t]
    \centering
    \begin{subfigure}[t]{0.7\columnwidth}
        \includegraphics[width = \columnwidth]{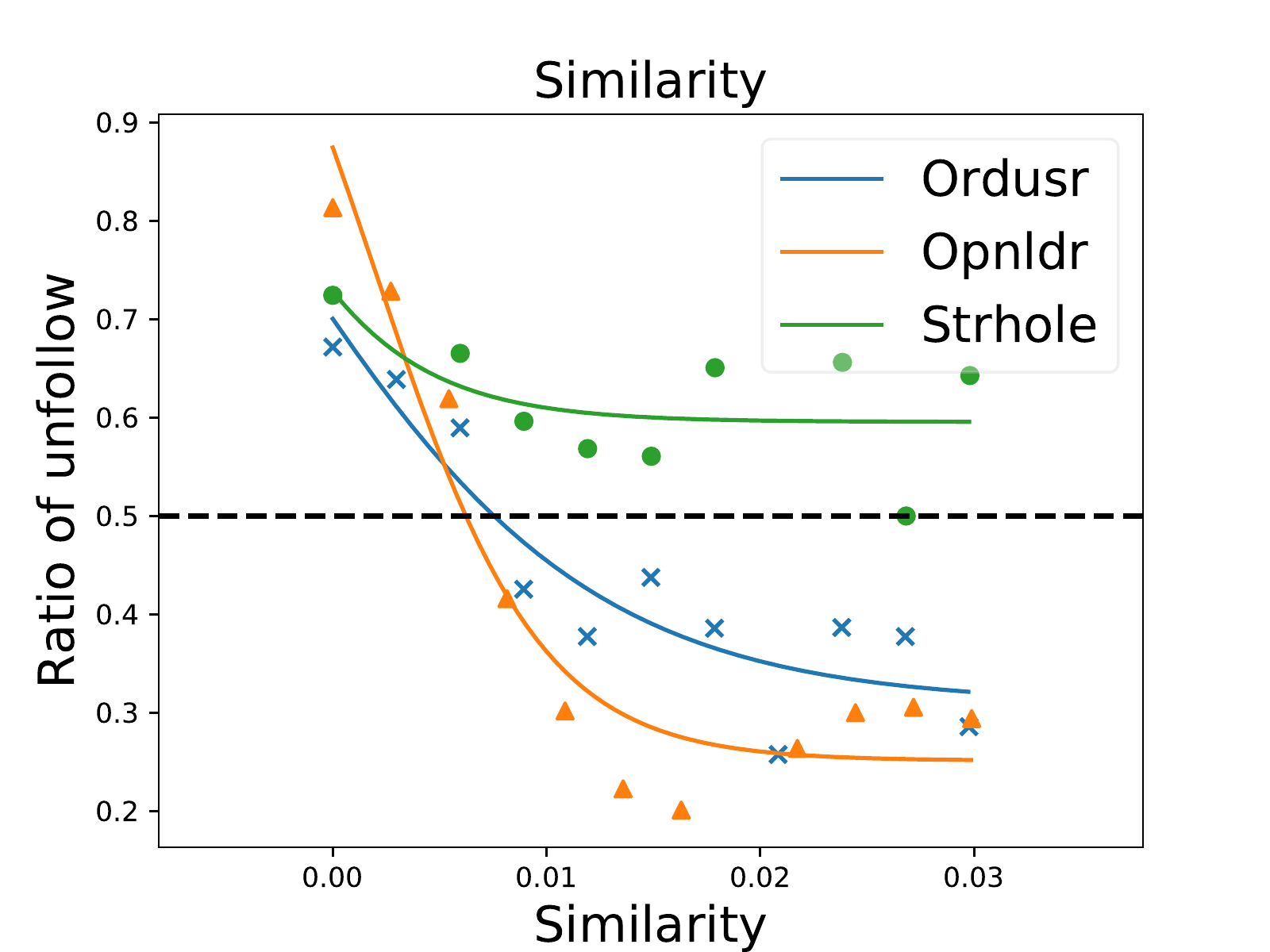}
        \caption{}
        \label{fig:similar}
    \end{subfigure}
    \begin{subfigure}[t]{0.7\columnwidth}
        \includegraphics[width=\columnwidth]{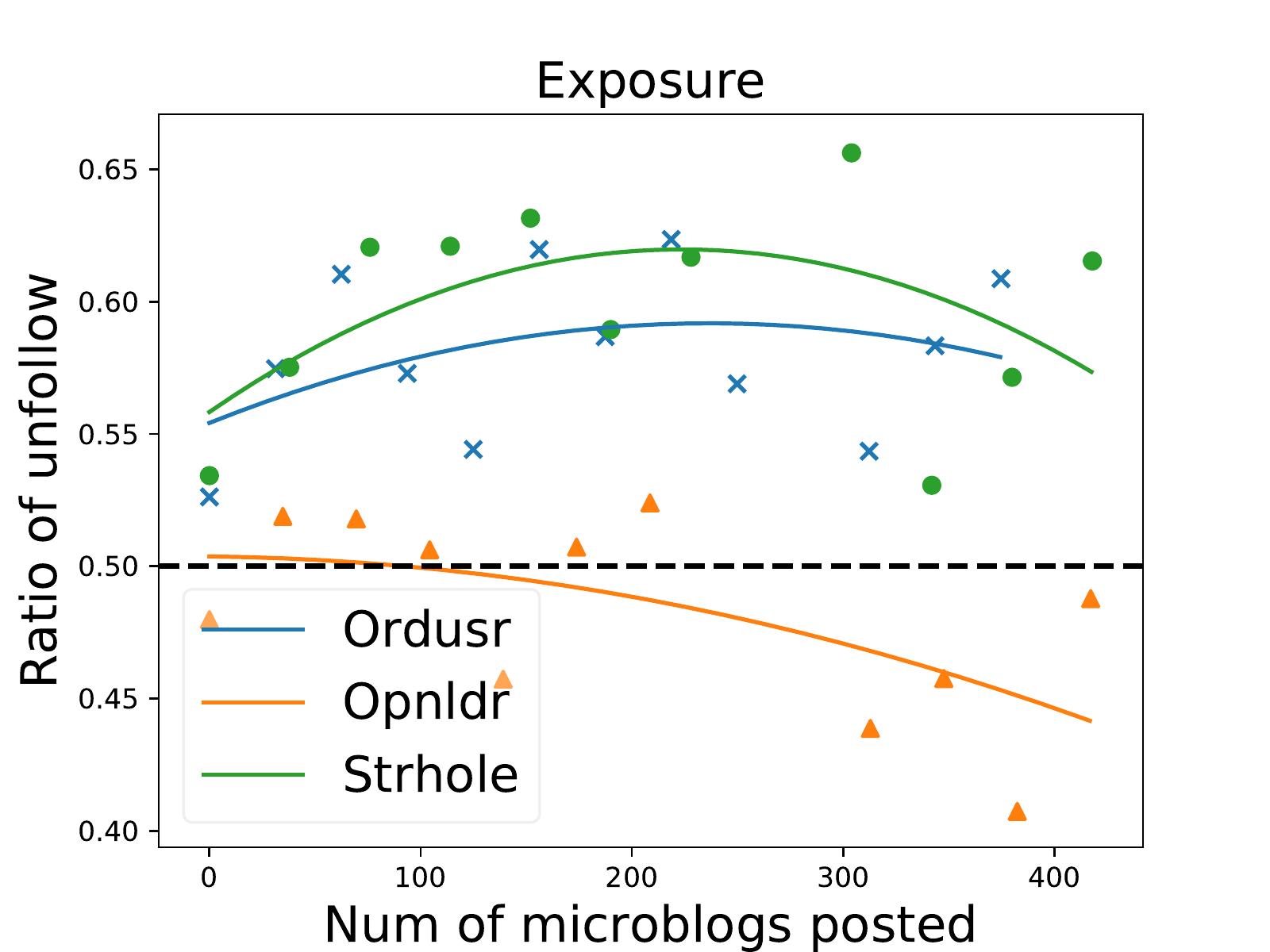}
        \caption{}
        \label{fig:expose}
    \end{subfigure}
    \caption{The results of exploratory analysis, (a) (b) respectively plots the distribution of $rou$ under the interaction between similarity / exposure and social role.}
\end{figure}

Although the unfollow behavior is pervasive, the ratio between unfollow relationships and hold relationships is still unbalanced~(2.53\% in our dataset). %
Therefore, we build a balanced sub-dataset $E_{test}$ for fair data observation and further training-test scheme. %
Table~\ref{tab:dataset_compose} shows the composition of $E_{test}$, %
here we filter out the edges in which either follower or followee post no microblog content. %

To measure how the interaction effects between spatial and temporal attributes affect user's decision on unfollow, %
we define metric $rou(c)$~(ratio of unfollow under condition $c$) %
to be the ratio bewteen the number of unfollow edges and the number of all edges. Formally: %
\begin{equation}
    rou(c) = \frac{\mathbf{N}_{un}(c)}{\mathbf{N}_{ho}(c) + \mathbf{N}_{un}(c)},
\end{equation}
where $\mathbf{N}_{un}(c)$ is the number of edges that break relationships under condition $c$ within the observed month, %
$\mathbf{N}_{ho}(c)$ is the number of edges that hold relationships. %
Edges with higher $rou(c)$ are more likely to break, while with lower $rou(c)$ are more stable. %
In the following analysis, we leverage $E_{test}$ to estimate the distribution of $rou(c)$ under different combinations of user's spatial attributes and temporal attributes. %

\textbf{Interaction between similarity and social role}. %
Similarity reveals user's homogeneity in short-term post content. Two users with higher homogeneity are more likely to have stable relationship. %
Figure~\ref{fig:similar} plots the relationship between similarity and $rou$ under different social roles. %
We observe that as similarity increases, the $rou$ quickly decreases to the saturation point. %
Figure~\ref{fig:similar} also shows that the interaction effects bewteen social role and similarity is significant: %
opinion leader is most sensitive to similarity, since it varies the most as similarity increases. %
On the other hand, for structure holes, similarity seems to be a weak factor. %

\textbf{Interaction between exposure and social role}. %
Exposure evaluates the activeness of followee in recent time. %
From Figure~\ref{fig:expose}, we observe that with the increase of exposure, $rou$ shows different tendency under different social roles. %
The $rou$ of opinion leaders is naturally lower than that of ordinary users and structure holes. %
Meanwhile, the $rou$ of ordinary users and structure holes fluctuates with the increase of exposure, %
while the $rou$ of opinion leaders monotonically decreases. Such phenomenon has a simple interpretation: followers are concerned with the messages from opinion leaders, %
and would not curtly break up relationships with them. %

\textbf{Summary}. We reveal that the spatial and temporal attributes interplay with each other and result in an intricate mechanism of unfollow behavior. %
Therefore, for unfollow prediction, it is necessary to employ the nonlinearity of neural network to model such interactions. %
In next section we will introduce how the UMHI model leverages neural network to extract discriminative features for unfollow prediction. %

\section{Proposed Model}
Based on the exploratory analysis in Section~\ref{sec:data_obs}, we propose a novel UMHI model which incorporates heterogeneous information to predict the unfollow behavior. %
Our model simultaneously takes the spatial attributes and temporal attributes as input. More specifically, (1) we capture spatial attributes through social network structure, %
and utilize network embedding to compress the graph structured data into feature vectors. (2) We infer user's temporal attributes from user-posted content and unfollow history, %
the hierarchical attention network~(HAN) and matrix factorization~(MF) are respectively leveraged to learn the feature vectors from post content and unfollow history. %
(3) We employ the nonlinearity of MLP layers to model the interaction effects between spatial attributes and temporal attributes. %
The overall architecture of UMHI is presented in Figure~\ref{fig:workflow}. %

The remainder of this section is organized to elaborate each component of UMHI framework. %

\subsection{Network Structure Encodes}
The user-located social network structure is closely related to the unfollow behavior. As revealed in Section~\ref{sec:data_obs},
users with higher pagerank score have lower probability to be unfollowed. Except for the social role which affects followee's decision on unfollow, %
prior work by Quericia \textit{et al.}~(2012b) argues that common friends between two users also greatly affects the stability of relationship. %

To comprehensively encode network structure into low dimensional feature vectors, we leverage the network embedding method. %
Different network embedding methods have different capacities for encoding different kinds of structure information~\cite{dalmia2018towards}. %
Because of different optimization strategies, LINE algorithm has better capacity on capturing local information, %
while Deepwalk and Node2vec prefer to encode global information. %

For unfollow prediction, experiment results in Section~\ref{sec:exp} demonstrate that local information plays a more important role than global. %
Therefore, we employ the LINE algorithm to represent social network structure. %
LINE has two different objectives: first-order proximity and second-order proximity. %
The first-order proximity refers to the local pairwise proximity between vertices in the network, %
while the second-order proximity is the similarity in the neighborhood network structures between two nodes. %
We denote LINE with first-order proximity as LINE1, and LINE with second-order proximity as LINE2. %
The UMHI framework incorporates both LINE1 and LINE2 as input, here we respectively denote the node embeddings of follower $i$ and followee $j$ as $\mathbf{n}_{i}$ and $\mathbf{n}_{j}$.

\subsection{Post Content Encodes}

\begin{figure}[!t]
    \centering
    \includegraphics[width=0.82\columnwidth]{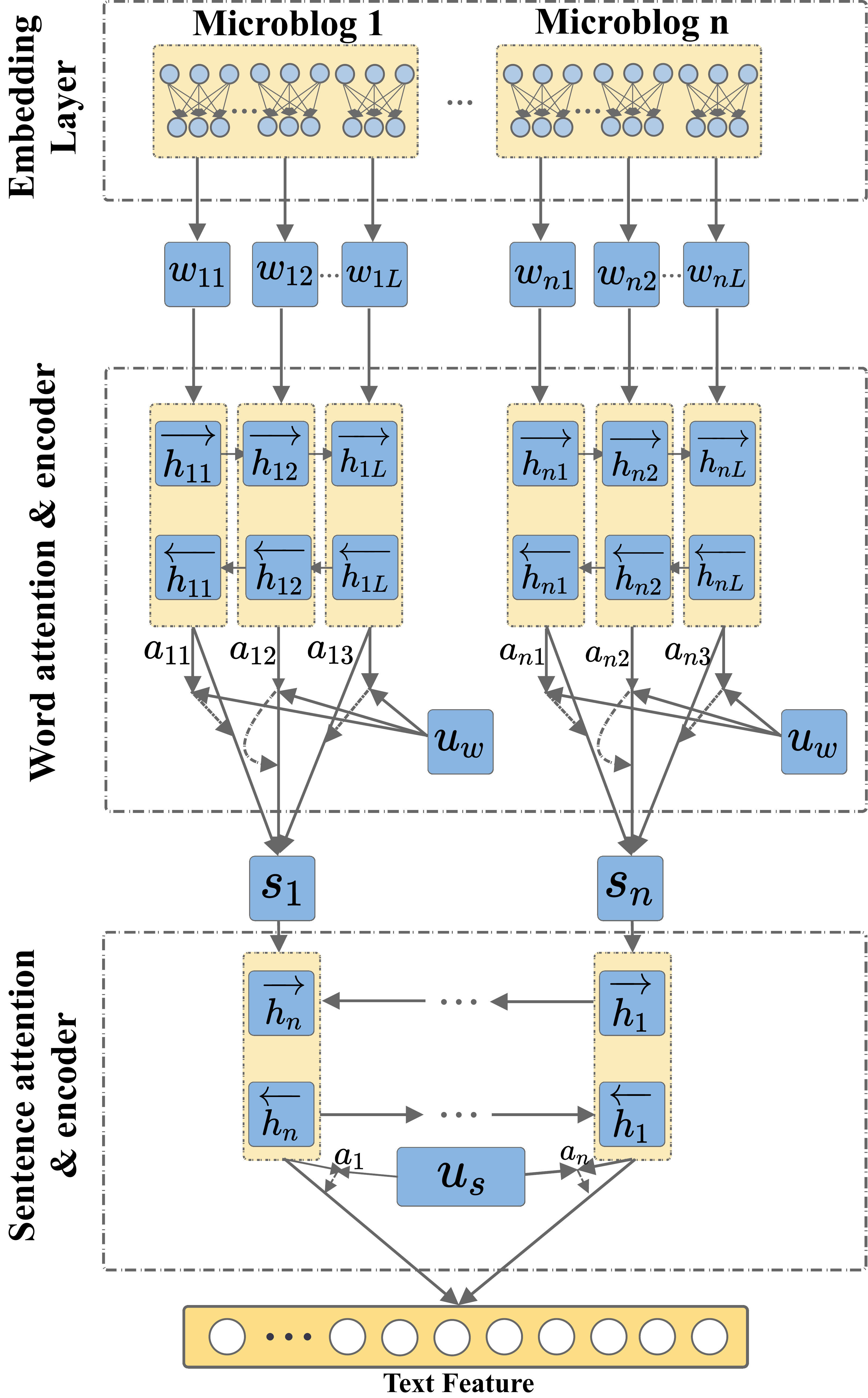}
    \caption{An illustration of hierarchical attention network. The first layer of HAN incorporates the word embeddings of each post as input, %
    then outputs a dimension fixed post content represention, the second layer of HAN incorporates post content representations of all posts as input, %
    and aggregates representations into a dimension fixed feature vector. }
    \label{fig:han}
\end{figure}

Users' post content reflects their temporal preferences. By observing the post content of follower and followee, %
we may guess whether the follower losts interest in the followee's posts at present. %
Since that both the words in each post and all posts of one user are serialized sequences, %
we employ the hierarchical attention network~(HAN) to extract the discriminative features. %
The hierarchical attention network, as shown in Figure~\ref{fig:han}, contains two bidirectional attention LSTM layers. %
The first LSTM layer encodes each post into dimension fixed representation, the second layer aggregates representations from all posts into a dimension fixed feature vector. %
Finally the attention mechanism~\cite{bahdanau2014neural} assign different weights to different words and posts. 

More specifically, we firstly embed each word of post content into word embeddings $w_{ij} \in \mathbb{R}^{N}$~($i$ is microblog ID, $j$ is word ID). %
For a microblog with $T$ words, the bidirectional attention LSTM $f_{1}$ takes $w_{i1} ... w_{iT}$ as input, and outputs the sentence representation $s_{i}$, formally as: 
\begin{equation}
    s_{i} = f_{1}(w_{i1} ... w_{iT}).
\end{equation}

After obtaining representation $s_{i}$ of each post, for one user with $L$ microblogs, %
the second LSTM layer $f_{2}$ aggregates the learned representation into a dimension fixed feature vector $\mathbf{m}$, formally:
\begin{equation}
    \mathbf{m} = f_{2}(s_{1} ... s_{L}).
\end{equation}

The overall UMHI framework simultaneously extracts the post content feature vector of follower $i$ and followee $j$, we respectively denote the two feature vectors as $\mathbf{m}_{i}$ and $\mathbf{m}_{j}$.

\subsection{Unfollow History Matrix Factorization}
Unfollow history records user's unfollowed-people list in recent time. Users with similar unfollow history show homogeneity in unfollow behavior. %
Specifically, if two users unfollowed similar followees recently, they may make similar unfollow decisions in the future. %
However, the unfollow behavior is sparse, statistics in our constructed dataset show that each user would only unfollow 6 users on average in a month. %
Therefore, it is hard to directly predict unfollow action by a user's unfollowed-people list. 

Inspired by the matrix factorization~(MF) based collaborative filtering algorithm, we construct a unfollow history matrix $R_{train}$ as defined in Section~\ref{sec:def}. %
By matrix factorization we compress the unfollow history of follower $i$ to latent vector $p_{i}$, %
and also map the unfollowed history of followee $j$ to latent vector $q_{j}$. The dot product of $p_i$ and $q_j$ produces $\hat{r}_{ij}$, %
which estimates the value of $r_{ij}$ in $R_{train}$. Formally: %
\begin{equation}
    \hat{r}_{ij} = p_{i}q_{j}^{T},
\end{equation}
where $p_i$ and $q_j$ are learned by optimizing the following regularized squared error:
\begin{equation}
    \min_{p^*,q^*}\sum_{\langle i,j \rangle \in |V| \times |V|} (r_{ij}-p_iq_j^T)^2+\lambda(||p_i||^2+||q_j||^2).
\end{equation}
The regularized squared error is minimized by stochastic gradient method~(SGD) without negative sampling. %
The UHMI framework then incorporates both $p_i$ and $q_j$ as input. %
Experiment results demonstrate that feature vectors $p_i$ and $q_j$ are effective for the final prediction. %

\subsection{Output and Objective Function}
Until now, given follower $i$ and followee $j$, we've obtained network embeddings $\mathbf{n}_i, \mathbf{n}_j$, %
post content feature vectors $\mathbf{m}_i, \mathbf{m}_j$, and unfollow history feature vectors $p_i, q_j$. %
Among which $\mathbf{n}_i$ and $\mathbf{n}_j$ are spatial attributes, while $\mathbf{m}_i, \mathbf{m}_j, p_i$ and $q_j$ are temporal attributes. %
To model the complicated interactions between spatial attributes and temporal attributes, %
we employ MLP layers to incorporate the unified representation as input. %
The nonlinearity of MLP layers can adequately represent the interaction mechanism between spatial attributes and temporal attributes. %
Finally, the output layer estimates the unfollow probability $y_{ij}$ between follower $i$ and followee $j$. Formally: %

\begin{equation}
    \mathbf{d} = \mathbf{n}_i \oplus \mathbf{n}_j \oplus \mathbf{m}_i \oplus \mathbf{m}_j \oplus p_i \oplus q_j,
\end{equation}

\begin{equation}
    y_{ij} = \mbox{Sigmoid}(\mbox{MLP}(\mathbf{d})).
\end{equation}
The objective function to be minimized is defined as:
\begin{equation}
    O = \sum_{\langle i,j \rangle \in |V| \times |V|} (r_{ij} \log (y_{ij}) + (1 - r_{ij}) \log (1 - y_{ij})),
\end{equation}
where $r_{ij}$ is the element of $R_{train}$. The objective function is a format of cross-entrophy. %
Since positive instances and negative instances are unbalanced, we sample equal number of positive and negative instances in each mini-batch.

\section{Experiments}
\label{sec:exp}
In this section, we firstly introduce dataset construction, evaluation metrics and UMHI implementation details, %
then we demonstrate the effectiveness of UMHI framework through comprehensive experiments. %
We show quantitative results on unfollow prediction, %
and respectively analyze spatial attributes, temporal attributes and the interaction effects through experimental results. %
Finally, we confirm the robustness of UMHI framework under different train-test split. %

\subsection{Dataset}
Since prior researches did not publish their datasets, we only conduct experiments on the dataset we built. %
Section~\ref{sec:data_obs} has introduced how we construct the sub dataset $E_{test}$. %
To evaluate the model performance and prevent information leakage, we conduct five-fold cross validation on $E_{test}$. %
Specifically, we randomly holdout 20\% of $E_{test}$ for test, and combine the remaining data with $R_{train} = R \backslash E_{test}$ for training. %

\subsection{Metrics}
We adopt the following three popular metrics to evaluate the performance of unfollow prediction:
\begin{itemize}
    \item \textbf{Precision}: It measures the probability that a predicted positive instance would be the true positive. %
    \item \textbf{Recall}: It measures the probability that the true positive would be predicted to be positive instance. %
    \item \textbf{AUC}: It measures the probability that a classifier will rank a randomly chosen %
    positive instance higher than a randomly chosen negative one.
\end{itemize}

\subsection{Implementation Details}
We first use jieba~\footnote{https://github.com/fxsjy/jieba} to cut microblog posts into separated words, %
use gensim~\footnote{https://radimrehurek.com/gensim/}'s word2vec model to embed words into embeddings, %
and implement our UMHI model with Keras. %

There are two stages for training our UMHI framework. In the first stage, we pretrain each component of UMHI, %
and in the second stage, we combine the three components of UMHI and fine tune the fusion MLP layer. %

\subsubsection{Stage \rom{1}: Pretrain of Each Component.}
During the pretrain stage, we respectively pretrain LINE, HAN, and Matrix Factorization. %
We set the embedding size of LINE to 100, and train LINE for 100 epochs. %
When pretraining HAN, we set the size of LSTM cell to 100 and train for 10 epochs. %
The network is optimized by the adam optimizer~\cite{kingma2014adam}, with the learning rate 0.001, and $\beta_1 = 0.1, \beta_2 = 0.001$. %
We optimize Matrix Factorization with latent size of 64, learning rate of 0.01, and 100 epochs. %

\subsubsection{Stage \rom{2}: Global Fine-Tuning.}
In global fine-tuning stage, we fix the parameters of LINE, HAN and Matrix Factorization, %
and train the MLP layers with adam optimizer for 10 epochs, the learning rate is set to 0.001. %
We choose the performance of models when the precision, recall and AUC of test set achieves biggest value during training. %

\subsection{Comparison Methods}
To justify the effectiveness of the proposed model, we compare the performance of our model with two kinds of baselines. %
Firstly, we compare UMHI with prior unfollow prediction methods. Prior unfollow prediction methods usually extract rule based features, %
and conduct training-test scheme on dataset with limited size, %
therefore showing inferior performance in our large-scale real-world setting. %
Secondly, we compare LINE with other network embedding methods so as to verify that local structure information plays an more important role than global information for unfollow prediction. %
The compared methods are listed as follows:
\begin{itemize}
    \setlength{\topmargin}{0pt}
    \setlength{\itemsep}{0em}
    \setlength{\parskip}{0pt}
    \setlength{\parsep}{0pt}
    \item \noindent \textbf{Doc2vec \& Action Features+LR~(DA + LR).} Referring to Maity's method \cite{maity2018did}, %
    we extract Doc2vec features and action features, then predict edge status by logistic regression.

    \item \noindent \textbf{Structural \& Action Features + LR~(SA + LR).} Referring to Kwak's method \cite{kwak2012more}, %
    we extract structural features and action features of users, then predict edge status by logistic regression.

    \item \noindent \textbf{Deepwalk.} Deepwalk~\cite{perozzi2014deepwalk} is a network embedding method, %
    which leverages truncated random walks to obtain the structural information of each vertex. %

    \item \noindent \textbf{Node2vec.} Node2vec~\cite{grover2016node2vec} is a network embedding method which designs a biased random walk procedure. %
    In experiment, we use grid search to choose random walk strategies with the best performance. %

    \item \noindent \textbf{The Proposed Method.} To demonstrate the effectiveness of different parts in UMHI model, %
    we assemble LINE1, LINE2, HAN and matrix factorization~(MF) into different combinations and compare their performance. %
\end{itemize}

\subsection{Experimental Results}

\begin{table}[htbp]
    \centering
    \caption{Results of unfollow prediction}
    \begin{tabular}{lrrr}
    \toprule
    Method & Precision & Recall & AUC\\
    \midrule
    DA + LR& 0.6292 & 0.5786 & 0.6840\\
    SA + LR & 0.6348 & 0.5270 & 0.6932\\
    Deepwalk & 0.6281 & 0.8123 & 0.6755\\
    Node2vec & 0.6288 & 0.8114 & 0.6707\\
    \midrule
    LINE1 & 0.6316 & 0.7893 & 0.6845\\
    LINE2& 0.6413 & 0.7553 & 0.6924\\
    LINE1 + LINE2 & 0.6458 & 0.7561 & 0.7059\\
    HAN & 0.6766 & 0.8292 & 0.7441\\
    MF & 0.7701 & 0.8686 & 0.8136\\
    LINE1 + LINE2 + HAN & 0.7034 & 0.8516 & 0.7718\\
    UMHI & \textbf{0.7868} & 0.8131 & \textbf{0.8673}\\
    \bottomrule
    \end{tabular}
    \label{tab:result}
\end{table}

Table~\ref{tab:result} displays the performance across different models, from this table, we have the following analysis. %

\begin{figure}[!t]
    \centering
    \begin{subfigure}[t]{0.6\columnwidth}
        \centering
        \includegraphics[width = \columnwidth]{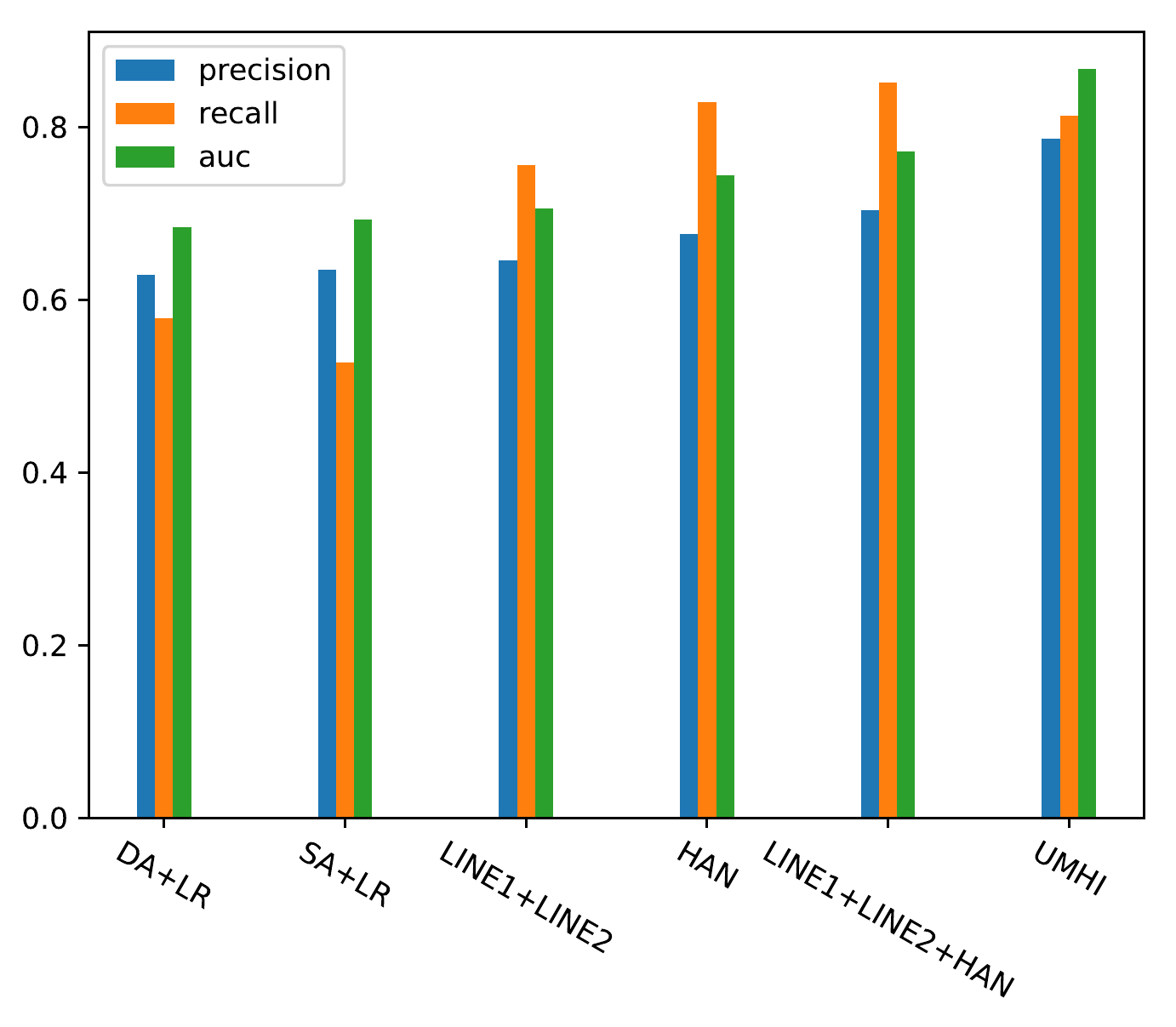}
        \caption{Precision, AUC and Recall of different comparision methods.}
        \label{fig:comparision}
    \end{subfigure}
    \centering
    \begin{subfigure}[t]{0.6\columnwidth}
        \centering
        \includegraphics[width = \columnwidth]{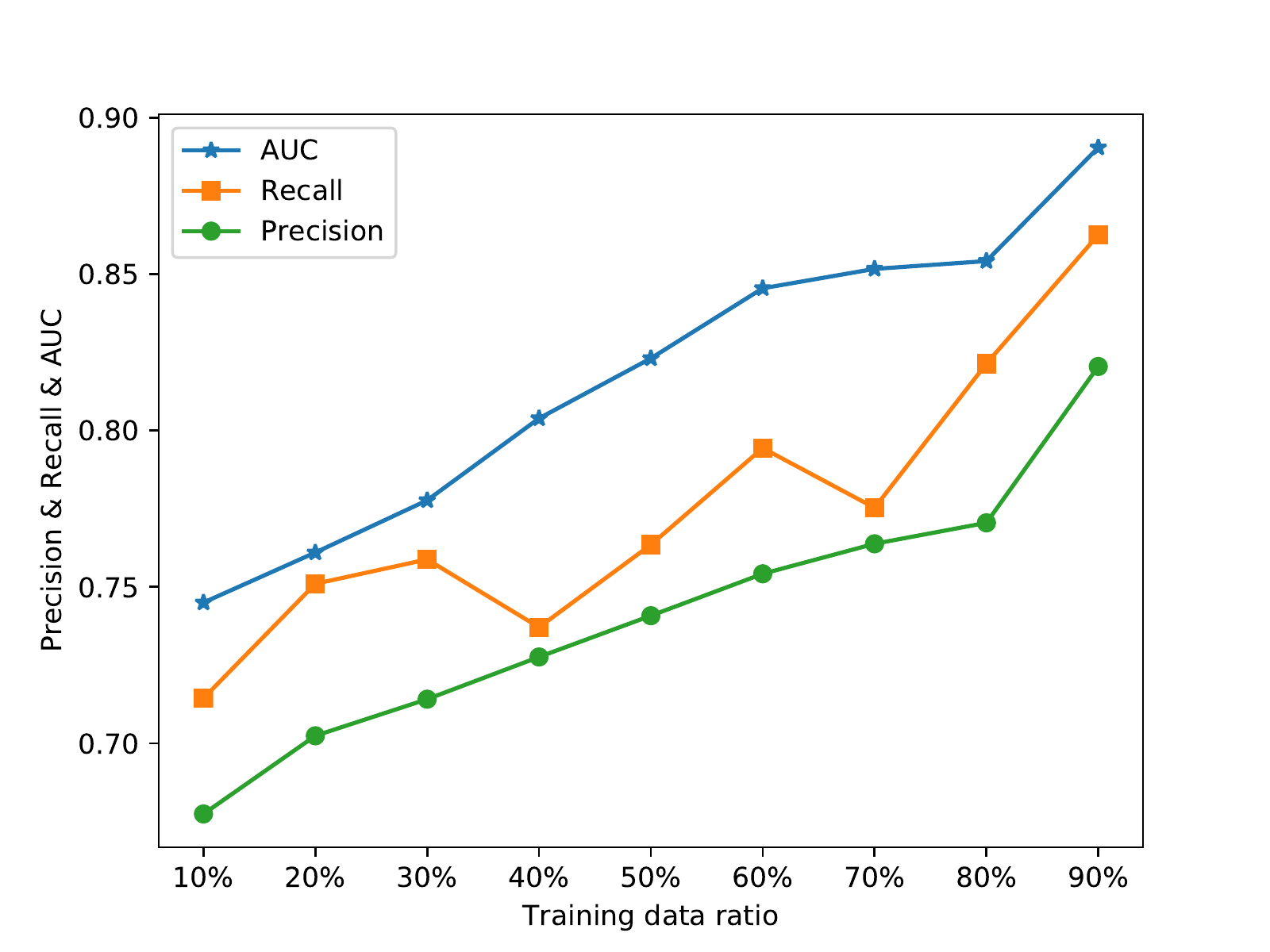}
        \caption{Precision and AUC with different training and test data size.}
        \label{fig:robustness}
    \end{subfigure}
    \caption{Experimental Results}
\end{figure}

\textbf{Overall Performance.} 
To verify the validness of interaction effects between spatial and temporal attributes, we make the following comparisons, as shown in Figure~\ref{fig:comparision}. %
Firstly, for the three methods~(DA + LR, SA + LR, and LINE1 + LINE2 + HAN) that take same sources of input, %
DA + LR and SA + LR are two handcrafted methods that can hardly represent interaction effects, %
while LINE1 + LINE2 + HAN can represent highly nonlinear interaction mechanism. %
Experiment results show that LINE1 + LINE2 + HAN outperforms DA + LR by 7.42\% and outperforms SA + LR by 6.86\% in terms of precision, %
verifying that the interaction effects are powerful. %

Meanwhile, from the comparison among LINE1 + LINE2, HAN, and LINE1 + LINE2 + HAN, %
we can observe that LINE1 + LINE2 + HAN significantly outperforms the first two models, confirming the effectiveness of interaction. %
Additionally, the comparison between LINE1 + LINE2 + HAN and UMHI demonstrate that feeding predictive unfollow history information would further boost prediction performance. %
Also, we notice that UMHI compromises on the recall value, that's because unfollow is a sparse behavior and UMHI tends to be conservative. %

\textbf{Spatial Attributes Comparison.} 
Compared with Deepwalk and Node2vec, we observe that LINE1 + LINE2 achieves an improvement of 1.73\% in terms of precision. %
Such improvement verifies that local structure is more important than global structure under the unfollow prediction setting. %
Meanwhile, we discover that LINE2 is notably better than LINE1 for unfollow prediction. %
This is because LINE1 only considers the relationship between two nodes, while LINE2 considers the common neighbors between the two nodes. %
Therefore it reveals that the shared environment of two users can reflect the strength of the relationship more accurately than the relationship itself.

\textbf{Temporal Attributes Comparison.} 
We compare the two temporal attributes: unfollow history and post content. %
Experiment results show that MF performs 9.35\% better than HAN in terms of precision, %
therefore unfollow history contributes significantly more than the post content. %


\textbf{Robustness Analysis.} To verify the robustness of UMHI framework, we change the proportion of training set and test set and redo the experiments. %
Results in Figure~\ref{fig:robustness} show that the model is effective under limited training data size. Even with small size of training set~(10\%-30\%), %
our model can still have an acceptable and steady performance.
\linespread{1.2}
\section{Conclusion}
In this work, we constructed a large-scale social network dataset for unfollow behavior mining. %
Our dataset contains 1.8 million Chinese users and records relation dynamics of these users in a month. %
Based on the constructed dataset, we conducted extensive analyses on how users' spatial attributes and temporal attributes affect their decisions on follow, %
and revealed the interaction effects between these two categories of attributes. %
Then, we proposed the UMHI framework to learn users' spatial attributes and temporal attributes through their footprints in online social networks. %
The proposed framework outperforms baseline methods by a large margin, and the detailed factor analyses show that each component of UMHI is effective. %

For future researches, the constructed dataset still contains ample social dynamics that deserve further exploring. During the recorded month, %
some followers have launched a burst of unfollow behaviors, and some relationships have experienced several status alterations, %
detecting such anomalies in online social networks is beneficial for some downstream tasks like depression detection and rumor detection. %

\section{Acknowledgments}
This work is supported by National Key Research and Development Plan (2016YFB1001200), %
the state key program of the National Natural Science Foundation of China (NSFC) (No.61831022), %
Beijing Academy of Artificial Intelligence (BAAI), %
and National Natural and Science Foundation of China(61521002).

{
\bibliographystyle{named}
\bibliography{aaai20}
}

\end{document}